\documentclass[3p,times]{elsarticle}

\usepackage{ecrc}

\volume{00}

\firstpage{1}

\journalname{Nuclear Physics A}

\runauth{A. Angerami for the ATLAS Collaboration}

\jid{nupha}

\usepackage{amssymb}
\usepackage{amsmath}
\usepackage{verbatim}
\usepackage{atlasphysics}
\newcommand{\AuAu}{\mbox{Au+Au}}
\newcommand{\PbPb}{\mbox{Pb+Pb}}

\newcommand{\Nevt}{\mbox{$N_{\mathrm{evt}}$}}

\newcommand{\Njet}{\mbox{$N_{\mathrm{jet}}$}}

\newcommand{\sqrtsnn}{\mbox{$\sqrt{s_{\mathrm{NN}}}$}}

\newcommand{\Rcollcent}{\mbox{$R_{\mathrm{coll}}$}}

\newcommand{\RFive}{\mbox{$R = 0.5$}}
\newcommand{\RFour}{\mbox{$R = 0.4$}}

\newcommand{\RTwo}{\mbox{$R= 0.2$}}
\newcommand{\Rcp}{\mbox{$R_{\rm CP}$}}

\newcommand{\antikt}{\mbox{anti-\kt}}

\newcommand{\ETfcal}{\mbox{$\Sigma E_{\mathrm{T}}^{\mathrm{FCal}}$}}

\newcommand{\kt}{\mbox{$k_{t}$}}
\newcommand{\sumet}{\mbox{$\Sigma E_{\mathrm{T}}$}}

\newcommand{\ETtbyf}{\mbox{$E_{\mathrm{T}}^{3\times 4}$}}
\newcommand{\ETsbys}{\mbox{$E_{\mathrm{T}}^{7\times 7}$}}

\newcommand{\pthat}{\mbox{$\hat{p}_{\mathrm{T}}$}}

\newcommand{\diff}{\mathrm{d}}

\newcommand{\RcpRatio}{\mbox{$R_{\mathrm{CP}}^{R}/R_{\mathrm{CP}}^{\,0.2}$}}
\newcommand{\papertype}{Proceedings}
\usepackage[figuresright]{rotating}

\begin{document}

\begin{frontmatter}


\title{Measurements of jet quenching and heavy flavor production with the ATLAS detector}
\author{A. Angerami for the ATLAS Collaboration}
\address{Physics Department, Columbia University,\\
  538 West 120th Street, New York, NY 10027, USA}
\ead{angerami@cern.ch}
%%\address{}
%\fntext[label2]{for the ATLAS Collaboration}

\begin{abstract}
Measurements of jet suppression in \PbPb\ collisions by the ATLAS
Collaboration are reported. The production of inclusive jet yields as
a function of jet \pt, collision centrality and jet size parameter $R$
are measured and presented through the central-to-peripheral ratio,
\Rcp. Jets are found to be suppressed in central collisions relative
to peripheral collisions by approximately a factor of two. The
suppression is found to show almost no variation with jet \pt, and the
\Rcp\ is found to increase slighly with increasing $R$. Measurements
of heavy quark energy loss are also presented using muons from
semi-leptonic decays of heavy flavor hadrons. The maximal suppression
is observed to be $\Rcp\sim0.5$ and shows no significant depedence on the muon \pt.

\end{abstract}


\end{frontmatter}

%%\linenumbers

\section{Introduction}
Relativistic heavy ion collisions provide access to matter at the highest temperatures ever produced in the laboratory. In these collisions the produced medium consists of a system of deconfined quarks and gluons known as the quark-gluon plasma (QGP). Jets produced in these collisions must traverse the medium and may suffer energy loss or modification of their parton shower through interactions. This phenomenon, known as jet quenching, provides direct access to the QGP transport properties through the jet-medium interaction. One of the first proposed signatures of QGP formation was the energy loss of back-to-back jets traversing different in-medium path lengths~\cite{Bjorken:1982tu}. Experimental signatures of partonic energy loss were first observed at RHIC through the measurement of the suppression of single hadrons~\cite{Adler:2003qi,Adams:2003kv,Arsene:2003yk,Back:2004ra} and modification of di-hadron angular correlations~\cite{Adler:2002tq,phenix:2008cqb} . Recently the phenomenon of jet quenching was observed directly using fully reconstructed jets at the LHC~\cite{Aad:2010bu, Chatrchyan:1327643}. The measurement of highly asymmetric dijets in central \PbPb\ collisions can be interpreted as an experimental realization of Bjorken's picture, where the large asymmetry is attributed to dijet pairs where one jet is produced near the surface and suffers little modification while the balancing jet suffers significant energy loss in the medium. Furthermore, the lack of modification of the dijet angular correlation indicates that the quenching mechanism must allow for significant energy loss without distorting the jet's angle.

The asymmetry measurements are sensitive to differential energy loss- the energy loss of one jet relative to another. Additional information about the quenching mechanism by considering the inclusive energy loss of jets, which is a measure of the average, absolute energy loss. As the jet production spectrum is steeply falling, the systematic downward shift in the jets' \pt\ caused by the energy loss results in a reduction of the yield of jets at a given \pt, a phenomenon referred to as jet suppression. The suppression can be quantified by comparing the jet \pt\ spectrum in central and peripheral collisions. As central collisions involve a larger nuclear overlap in the collision, there is a geometric enhancement to the hard-scattering rates. To assess the effects of the quenching this enhancement must be removed, thus the per-event jet yields must be normalized by a factor, \Rcollcent, which corrects for the centrality-dependent geometric enhancement of parton luminosity per nuclear collision. The suppression is then evaluated in terms of a nuclear modification factor,
\begin{equation}
\Rcp(\pT)\Big|_{\mathrm{cent}} =
\dfrac{1}{\Rcollcent}
\dfrac{
\left.\dfrac{\Njet}{\Nevt}\,\right|_{\mathrm{cent}} 
}{
\left.\dfrac{\Njet}{\Nevt}\,\right|_{\mathrm{periph}} 
}\,.
\label{eq:rcpdef}
\end{equation}
This quantity has been measured first at RHIC~\cite{Adler:2003qi,Adams:2003kv,Arsene:2003yk,Back:2004ra} and later at the LHC~\cite{CMS:2012aa,Aamodt:2010jd} for single hadrons, however such measurements do not provide the full information about the quenching mechanism as they are only sensitive to part of the jet. At the LHC, measurements of fully reconstructed jets are possible due to both the larger acceptance of the detectors, especially calorimeters with large pseudorapidity converage, as well as the enhanced hard scattering rates for jets with energies significantly above the underlying event (UE) fluctuations. The clustering in jet algorithms such as the anti-\kt\ ~\cite{Cacciari:2008gp} are controlled by a size parameter $R$, which corresponds to a nominal cone radius of $R=\sqrt{\Delta\eta^2+\Delta\phi^2}$ about the jet axis. The angular pattern of medium-induced radiation can be investigated by studying the suppression of jets with different $R$ values, and can thus provide quantitative constraints on jet quenching calculations~\cite{Vitev:2008rz}.

Heavy flavor jets provide additional insight into the quenching mechanism; the additional scale introduced by the large quark mass may force a different interplay between elastic and inelastic energy loss than that for light quarks~\cite{Dokshitzer:2001zm}. However, measurements of heavy quark production at RHIC via semi-leptonic decays to electrons showed a combined charm and bottom suppression in \AuAu\ collisions comparable to that observed for inclusive hadron production~\cite{Adare:2006nq,Abelev:2006db,Adare:2010de}. There is disagreement in the theoretical literature regarding the interpretation of the RHIC heavy quark suppression measurements~\cite{Djordjevic:2011tm,Wicks:2005gt,Gossiaux:2008jv,Uphoff:2011ad}  regarding the role of non-perturbative effects~\cite{Adil:2006ra,vanHees:2007me,Horowitz:2008ig}. Heavy quark jets can be studied by considering the semi-leptonic decay process of heavy flavored hadrons within the jet. At LHC energies this process is the primary source of single muon production at intermediate \pt\ values (4-14~\GeV)~\cite{Aad:2011rr}. Thus the suppression of single muons provides an indirect measurement of the jet suppression of heavy quark jets.

In these \papertype\ the suppression of inclusive jets reconstructed with the anti-\kt\ algorithm using $R$ values, $R=0.2,\,0.3,\,0.4$ and $0.5$ and the jet \Rcp\ is presented as a function of \pt, $R$ and centrality, and the discussion closely follows the results of a recent ATLAS paper~\cite{jetsupp}. All results are fully unfolded for experimental effects using an unfolding procedure based on the singular value decomposition (SVD)~\cite{Hocker:1995kb}. As fully reconstructed jets in heavy ion collisions are a relatively new experimental tool, special attention is taken to discuss potential problems with these measurements and the measures taken to address these issues in the analysis. A measurement of the suppression of single inclusive muons is also presented. A template technique was used to perform a statistical separation of the prompt muon signal fraction from the total measured inclusive muon spectrum which includes experimental background from decays in flight of pions and kaons as well as muons produced due to interactions in the detector. 

\section{Expermental setup and data sample}
The ATLAS detector consists of three main systems constructed in a barrel/end-cap geometrical confguration~\cite{Aad:2008zzm}. The inner detector (ID) provides charged particle tracking over $|\eta| < 2.5$. It is composed of silicon pixel detectors in the innermost layers, followed by silicon microstrip detectors and a straw-tube tracker, all immersed in a 2~T axial magnetic field provided by a solenoid. The ATLAS calorimeter system is composed of both electromagnetic (EM) and hadronic sampling calorimeters providing multiple longitudal sampling layers. These include the EM barrel and end-cap liquid Argon (LAr) calorimeters ($|\eta| < 3.2$), steel-scintillator hadronic tile calorimeter ($|\eta| < 1.7$), a LAr hadronic end-cap ($1.5 < |\eta| < 3.2$). The range is extended by the LAr forward calorimeters (FCal) $3.2 < |\eta| < 4.9$ which provides a combination of electromagnetic and hadronic energy measurements. Beyond the calorimeter is a muon spectrometer (MS) which tracks muons in the range $|\eta| < 2.7$, and is composed of additional tracking chambers in a magnetic field produced by three air-core toroid systems. The Zero Degree Calorimeters (ZDC), positioned at $|\eta| > 8.3$, provide measurements of spectator neutrons and were used to define a trigger requiring a pulse height above a threshold set below the single neutron peak. The Minimum Bias Trigger Scintillators (MBTS) provide measurements of charged particles in the range $2.1 < |\eta| < 3.9$ and consists of two pairs of counters placed at $z = \pm 3.6$~m.

The measurements presented here were performed using data from the 2010 $\sqrtsnn=2.76$~\TeV\ \PbPb\ run corresponding to a total integrated luminosity of $7\mu\mathrm{b}^{-1}$. A minimum bias sample was selected by requiring a ZDC coincidence trigger. In addition to the trigger requirements events were required to have a good reconstructed primary vertex and a measured time difference between the two sides of the MBTS of $\Delta t_{\mathrm{MBTS}} < 3$~ns. The combination of trigger and offline requirements select minimum-bias hadronic \PbPb\ collisions with an efficiency of $98 \pm 2\%$. The resulting data sample contained approximately 50 million events after applying these requirements. The centrality determination was performed by measuring the minimum bias total FCal \ET, \ETfcal, distribution and partitioning the range into percentile bins of the total integral. The event selection criteria and centrality definitions correspond to the standard ATLAS defintions for the analysis of 2010 data~\cite{ATLAS:2011ah}. For both the jet and muon suppression measurements the 60--80\% centrality bin was used as the peripheral reference in Eq.~\ref{eq:rcpdef}. 

Two separate Monte Carlo (MC) samples were used in assessing reconstruction performance and extracting corrections. The MC overlay sample was constructed by embedding PYTHIA~\cite{Sjostrand:1993yb} dijet events within minimum bias HIJING events~\cite{Wang:1991hta}. The separate samples of PYTHIA events were generated at different ranges of the \pthat\ in the hard-scattering process to insure good statistics over the full range of jet \pt. These samples were used for both the jet and muon suppression analyses. A separate sample of minimum bias HIJING events with no embedding of PYTHIA jets, the HIJING only, sample was produced in an analogous fashion for the purposes of studying the absolute fake rate and UE fluctuations.

\section{Jet reconstruction}
The jets used in this analysis were reconstructed using the anti-\kt\ algorithm with $R=0.2,\,0.3,\,0.4$ and 0.5. The effects of the UE were removed by applying a background subtraction procedure. Calorimeter towers with $\Delta \eta \times \Delta \phi\simeq 0.1\times 0.1$ were constructed from calorimeter cells and then clustered using the anti-\kt. The background was then subtracted on a per cell basis. For the $i^{\mathrm{th}}$ cell in a jet 
\begin{equation}
E_{\mathrm{T}\,i}^{\mathrm{sub}} = {\ET}_i -A_i\,\rho_j(\eta_i)\left(1 + 2v_{2\,j} \cos\left[2\left(\phi_i -\Psi_2\right)\right]\right)\,.
\end{equation}
The average density, $\rho$, and flow modulation, $v_2$, are determined separately for each calorimeter sampling layer $j$, and an iterative procedure is used to ensure the jets themselves do not bias the estimates of $\rho$ and $v_2$. The event plane angle $\Psi_2$ is determined from the azimuthal \ET\ distribution in the FCal, which is in a separate $\eta$ interval ($3.2 < |\eta| < 4.9$) from the jets considered here. Each cell is considered to be a four-vector using the $E_{\mathrm{T}\,i}^{\mathrm{sub}},\,\eta_i$ and $\phi_i$ with the mass taken to be zero. The jet's four-vector is then taken to be the four-vector sum of each of its constituent calorimeter cells. 

\begin{figure}[h]
\centering
\includegraphics[width=0.49\textwidth]{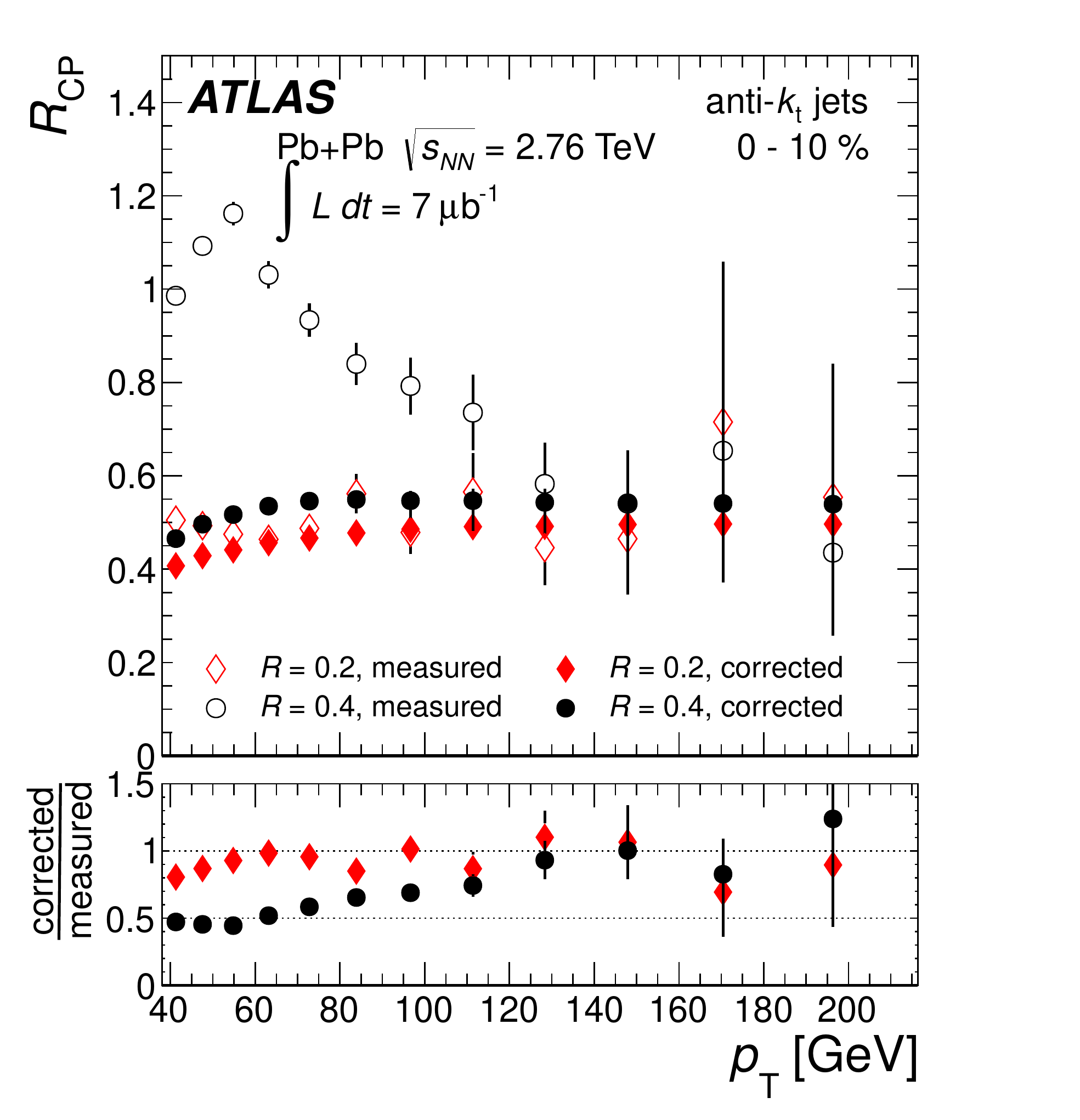}
\includegraphics[width=0.4\textwidth]{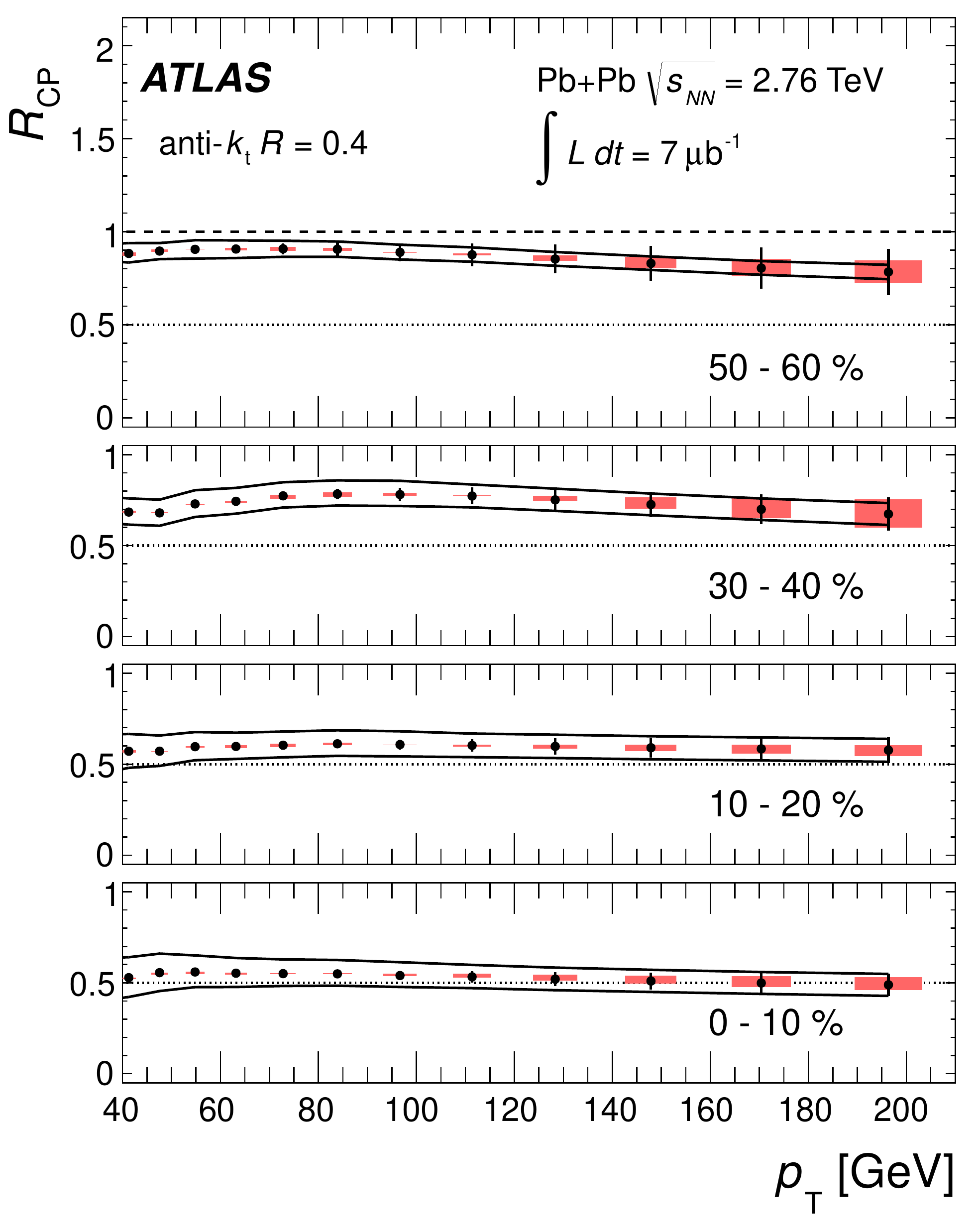}
\caption{
Left: measured and unfolded \Rcp\ values for the 0--10\% centrality
bin as a function of jet
\pT\ for \RFour\ and \RTwo\ jets with the ratio of 
unfolded to measured \Rcp\ values for both jet radii shown below. The error bars
on the points represent statistical uncertainties only.
Right: Unfolded \Rcp\ values as a function of jet
\pT\ for \RFour\ \antikt\ jets 
in four bins of collision centrality. The error bars indicate
statistical errors from the unfolding, the shaded boxes indicate
unfolding regularization systematic errors that are partially
correlated between points. The solid lines indicate 
systematic errors that are fully correlated between all points. 
The horizontal width of the systematic error band is chosen for
presentation purposes only. Dotted lines indicate $\Rcp =
0.5$, and the dashed lines on the top panels indicate $\Rcp = 1$~\protect\cite{jetsupp}.}
\label{fig:rcp_unfolding}
\end{figure}

The energy density of the UE has contributions from the mean background as well as global correlations and local uncorrelated fluctuations. The subtraction procedure above removes the contribution of the mean and also the most significant global correlation, the elliptic flow. However, uncorrelated $\ET$ fluctuations inside the nominal jet cone may significantly distort the measured jet energy on a jet-by-jet basis. This effect, combined with fluctuations in the energy deposition result in a jet energy resolution (JER) causing bin migration in the measured quantity. As the jet spectrum is steeply falling, the effect of the JER is to cause a net shift of the spectrum to higher \pt. This effect must be corrected for by an unfolding procedure, which requires a detailed knowledge of both the detector response and the UE fluctuations. The jet \Rcp\ as a function of \pt\ for the \RTwo\ (red diamond) and \RFour\ (black circle) jets is shown on the left panel of Fig.~\ref{fig:rcp_unfolding} both before (hollow markers) and after (solid markers) a correction for unfolding and efficiency, with the ratio shown below. As the UE fluctuations grow with area, the result for the smaller \RTwo\ jets changes little after the unfolding correction except at the lowest \pt. However, the difference for the \RFour\ jets is much greater and is nearly the magnitude of the measured signal, which emphasizes the requirement of having experimental control over the fluctuations to perform a meaningful measurement. Although the most technically challenging aspect of the measurement, the specifics of the unfolding procedure will not be discussed further in these Proceedings. For specific details of this procedure the reader is referred to Refs.~\cite{jetsupp,Angerami:2012tr}.

\section{Jet performance validation}
\begin{figure}[h]
\centering
\includegraphics[height= 4 in]{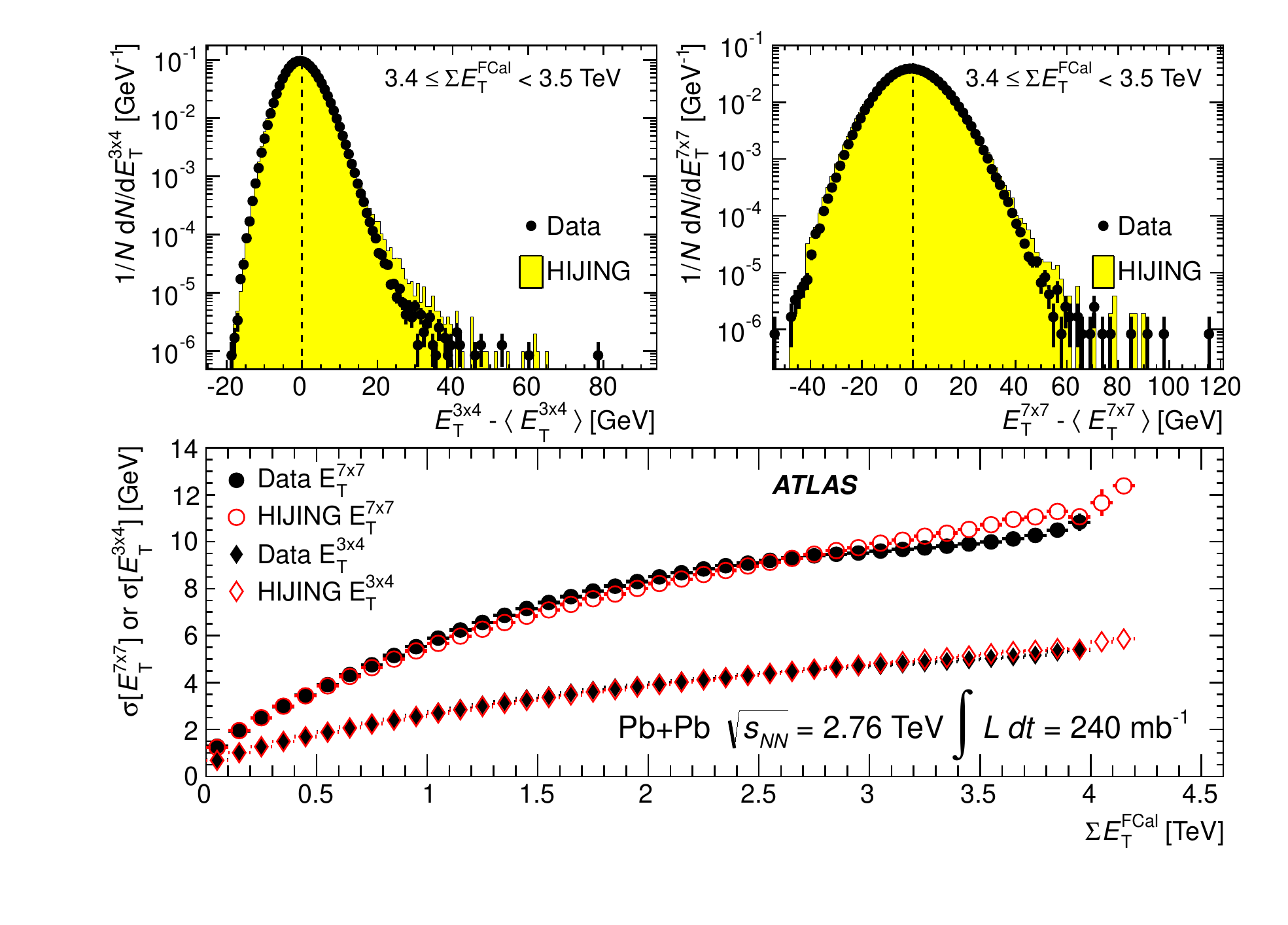}
\caption{Top: Representative distributions of $\ETtbyf - \langle\ETtbyf\rangle$
 (left) and $\ETsbys - \langle\ETsbys\rangle$ (right) for data (points) and MC (filled histogram) for
 \PbPb\ collisions with $3.4 \leq \ETfcal < 3.5$~\TeV. The vertical lines
 indicate $\ETtbyf - \langle\ETtbyf\rangle = 0$ and $\ETsbys -
 \langle\ETsbys\rangle = 0$.
 Bottom: Standard deviations of the \ETtbyf\ and
 \ETsbys\ distributions, $\sigma[\ETtbyf]$ and $\sigma[\ETsbys]$,
 respectively, in data and HIJING MC sample as a function of \ETfcal~\protect\cite{jetsupp}.}
\label{fig:fluct}
\end{figure}

A separate study was performed to provide a quantitative comparison between the UE fluctuations in the data and those present in HIJING, which was used to determine the jet performance and extract the response matrices used in the unfolding. In this fluctuations study the HIJING only MC sample was used. The total transverse energy, \sumet, in rectangular ``windows'' of adjacent calorimeter towers was computed for all windows on the range $|\eta| < 2.8$. The study was performed for different window sizes, where the sizes were chosen so correspond to the nominal area of anti-\kt\ jets (e.g. $3\times4 \Leftrightarrow $ \RTwo\ and $7\times7 \Leftrightarrow $ \RFour). The distribution of \sumet\ was measured in fine bins of \ETfcal, and is shown in Fig.~\ref{fig:fluct} for the $3\times4$, \ETtbyf, and $7\times7$, \ETsbys, window sizes in the top left and right panels respectively for both the data and the HIJING MC sample on the range $3.4\leq \ETfcal < 3.5$~\TeV, which lies in the 0-1\% centrality interval. In both cases the mean obtained from the data has been subtracted. The results indicate that shape of the fluctuations spectrum produced by HIJING and that present in the data agree extremely well over several orders of magnitude. The tail at the high end can be attributed to jets, which were not removed from the sample. The magnitude of the UE fluctuations that may be present underneath a jet after the average background is subtracted is given by the width of this distribution. For each bin in \ETfcal\ the standard deviation of these distributions was computed for both data and the MC sample, with the results shown in the bottom panel of Fig.~\ref{fig:fluct}. This result indicates that the UE fluctuations present in HIJING, which are used to provide the response matrices for the unfolding procedure, are consistent with the data over the full centrality range. The data-MC differences obtained from the fluctuations analysis were included in the systematic uncertainties in the jet spectrum measurement.

Large UE fluctuations may be falsely identified as jets, and the contribution of such ``fake'' jets must be understood and removed from the spectrum. This was achieved by requiring an additional signature of hard particle production in association with a reconstructed jet. For these purposes track jets, jets reconstructed from charged particles with $\pt > 4$~\GeV\ using the anti-\kt\ algorithm with \RFour, and electromagnetic clusters were used. Jets were required to be within $\Delta R < 0.2$ of either a track jet or electromagnetic cluster with $\pt > 7$~\GeV. This procedure was found to reduce the absolute rate of fake jets by a factor of approximately 50, while introducing only a slight additional \pt-dependent inefficiency which was subsumed into the correction for the overall reconstruction inefficiency. The residual fake rate after rejection was estimated to be 3\% of the signal rate for $\pt > 50$~\GeV.

\section{Jet suppression results}
\begin{figure}
\centerline{
\includegraphics[width=0.47\textwidth]{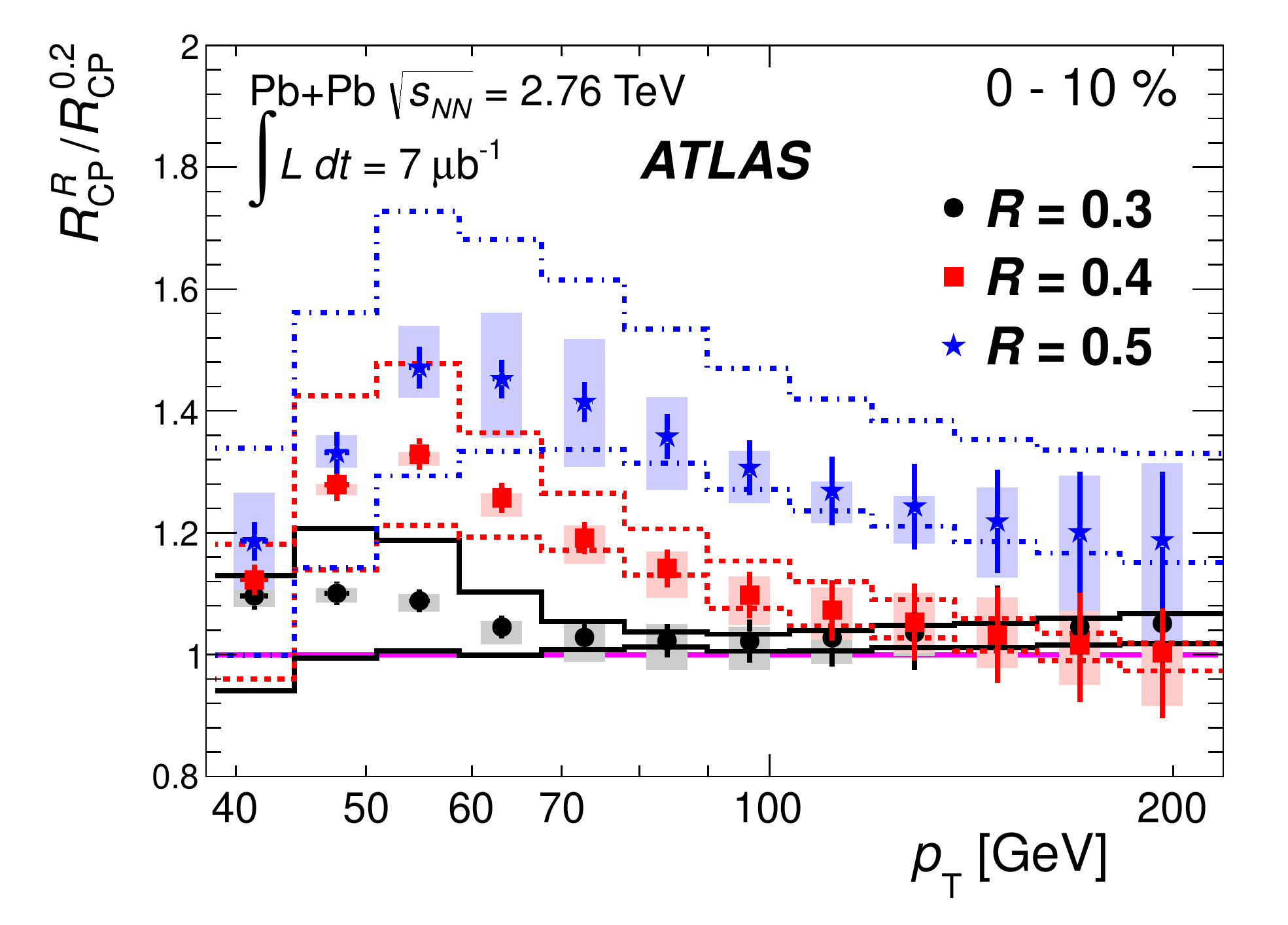}
\includegraphics[width=0.49\textwidth]{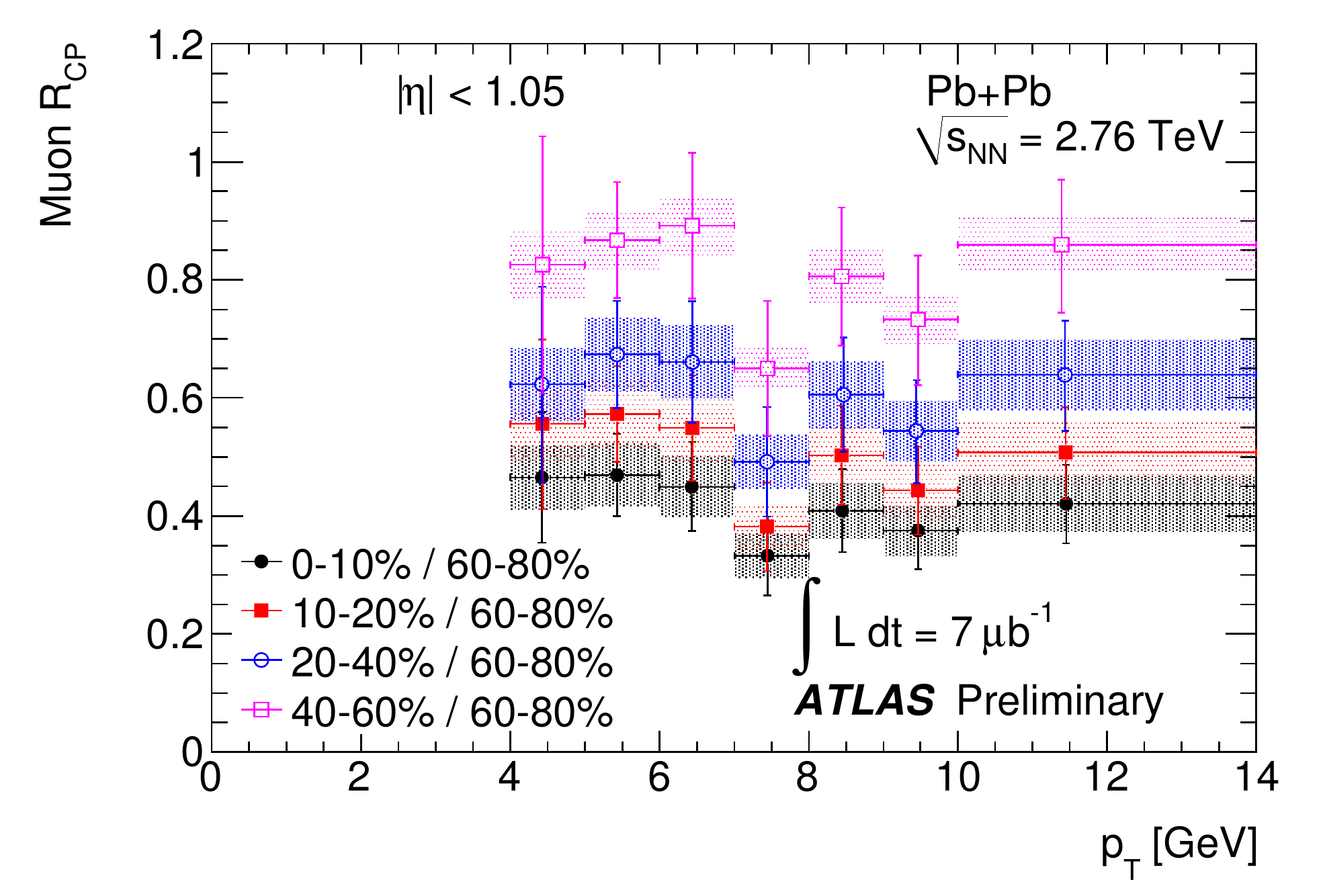}
}
\caption{
Left: ratios of \Rcp\ values between $R = 0.3, 0.4$ and 0.5 jets and $R =
0.2$ jets as a function of \pT\ in the 0--10\% centrality bin. The
error bars show statistical uncertainties. The shaded boxes
indicate partially correlated systematic errors. The lines indicate
systematic errors that are fully correlated between different \pT\ bins~\protect\cite{jetsupp}.
Right: muon \Rcp\ as a function of \pt\ in four centrality bins. The shaded boxes indicate systematic uncertainties that are fully correlated between \pt\ bins. The error bars indicate combined systematic and statistical errors, which are taken to be uncorrelated~\protect\cite{ATLAS-CONF-2012-050}.}
\label{fig:rcp_ratios}
\end{figure}

The unfolded and efficiency-corrected jet spectra were used to construct the \Rcp\ ratios defined in Eq.~\ref{eq:rcpdef} in bins of jet \pt, $R$ and centrality. The \Rcp\ for \RFour\ jets as a function of \pt\ is shown on the right side of Fig.~\ref{fig:rcp_unfolding} for four centrality bins. The black error bands indicate systematic uncertainties that are fully correlated among different \pt\ bins, while the shaded red boxes contain systematic uncertainties that are uncorrelated; the error bars indicate statistical uncertainties which develop a significant covariance through the unfolding. In the most central collisions, the \Rcp\ reaches a value of approximately $0.5$ and shows a weak \pt\ dependence. This \pt\ dependence is common to all centrality bins, with the overall level of suppression diminishing in increasingly peripheral collisions. The $R$ dependence of the suppression is shown in the left panel of Fig.~\ref{fig:rcp_ratios}, where the \Rcp\ ratio measured with a particular $R$ value, $R_{\mathrm{CP}}^{R}$, is divided by the corresponding value for \RTwo\ jets. As many of the uncertainties are correlated between the different $R$ values, many cancel in the ratio. The statistical error bars shown account for the full statistical correlation between jets of different $R$ values propagated through the unfolding procedure.
The results in that figure indicate a significant dependence of \Rcp\ on jet radius collisions. For $\pT < 100$~\GeV\, the \RcpRatio\ values for both \RFour\ and \RFive\ differ from one beyond the statistical and systematic uncertainties. This deviation persists for \RFive\ above 100~\GeV. A similar, but weaker dependence is observed in the 10--20\% centrality bin, and in more peripheral bins, no significant radial dependence is observed. The differences between \Rcp\ values for the different jet radii increase with decreasing \pT, except for the lowest two \pT\ bins. However, direct comparisons of \Rcp\ between different jet radii at low \pT\ should be treated with care as the same jets measured using smaller radii will tend to appear in lower \pT\ bins than when measured with a larger radius. 

\section{Inclusive muon analysis}
In the interval $4 < \pt < 14$~\GeV, the inclusive production of single muon is dominated by contributions from the semi-leptonic decays of hadrons containing open heavy flavor. Contributions from electroweak boson decays as well as Drell-Yan production and quarkonia decays all contribute at less than the 1\% level over this momentum range~\cite{Aad:2011rr}. Besides the heavy flavor signal, the inclusive muon yield contains backgrounds from $\pi$ and $K$ decays inside the detector as well as muons produced in hadronic showers in the calorimeter. These muons have different reconstruction properties, which were used to construct a discriminant. The templates were constructed from composite discriminant distributions, $\diff P/\diff C$, of both signal and background sources in an MC sample and the signal fraction was extracted using a statistical separation. A more detailed presentation of this analysis is presented in Ref.~\cite{ATLAS-CONF-2012-050}.

\begin{figure}[h]
\centering
\includegraphics[height=4.5 in]{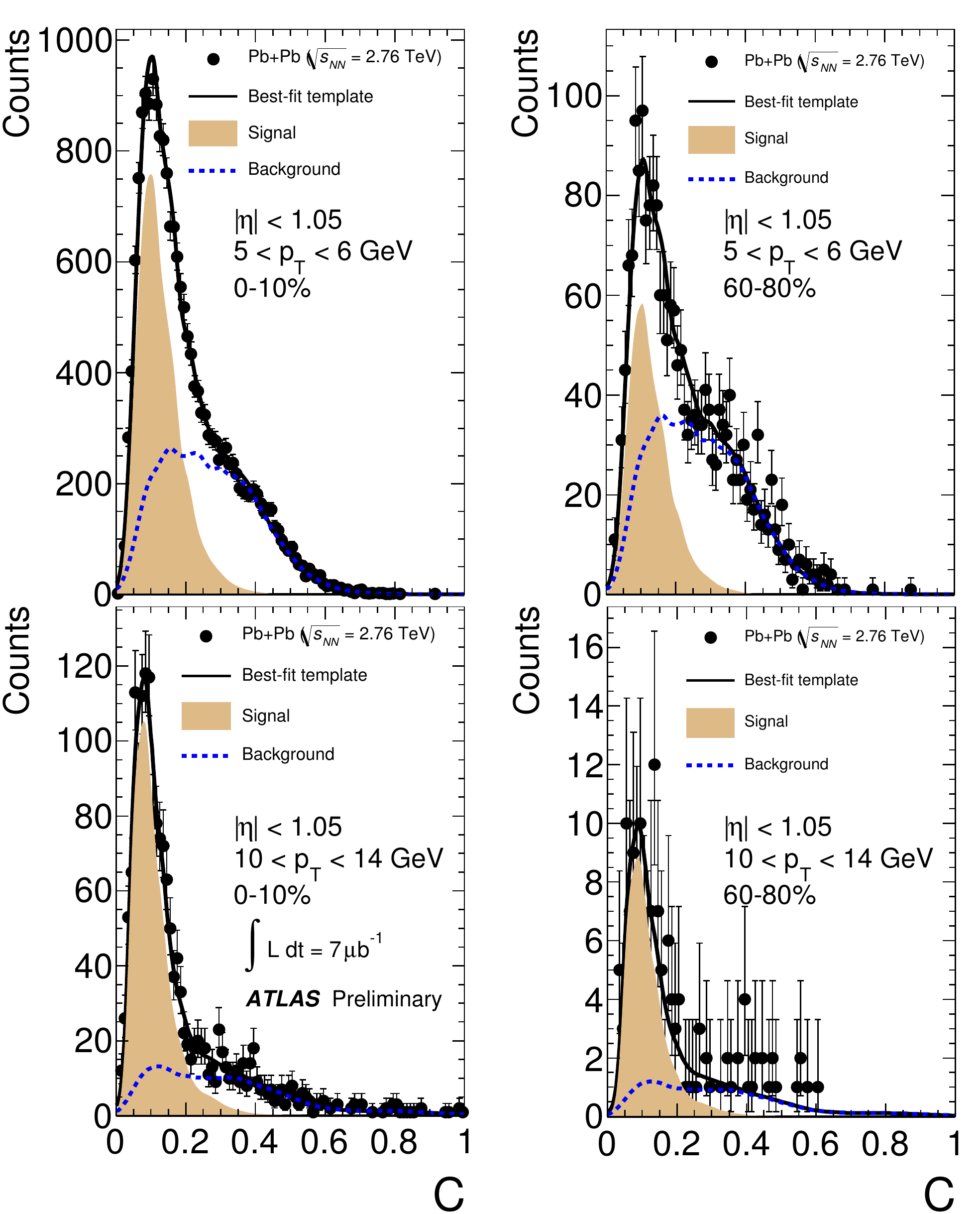}
\caption{Template fits for the composite discriminant distributions $\dfrac{\diff P}{\diff C}$, with $C$ defined in Eq.~\protect\ref{eq:composite_definition} for muons with  $5<\pt<6$~\GeV\ (top) and $10<\pt<14$~\GeV\ (bottom) in the 0--10\% (left) and 60--80\% (right) centrality bins. The black points indicate the data, with the black curve indicating the best fit to the two-component template. The signal and background contributions are indicated by the brown filled and blue dashed blue curves respectively~\protect\cite{ATLAS-CONF-2012-050}.}
\label{fig:muon_templates}
\end{figure}

Muons are reconstructed by combining information from the ID and MS as well as a parameterized form of the typical muon energy loss inside the calorimeter. For background muons, which are produced within the detector, the momentum measurements in the ID and MS will differ by more than the typical energy loss. Therefore the fractional energy loss distribution,
\begin{equation}
\dfrac{p_{\mathrm{loss}}}{p_{\mathrm{ID}}}=\dfrac{p_{\mathrm{ID}}-p_{\mathrm{MS}}-p_{\mathrm{param}}(p_{\mathrm{MS}},\eta,\phi)}{p_{\mathrm{ID}}}\,,
\end{equation}
will differ for signal and background. Furthermore, tracks corresponding to muons produced from decays in the ID may exhibit kinks at the decay vertex which will exceed the typical angular deflection due to multiple scattering, $\phi_{\mathrm{msc}}$. This can be quantified by considering the relative deflection at the $k^{\mathrm{th}}$ scattering center,
\begin{equation}
S(k)=\dfrac{1}{\sqrt{n}}\left(\displaystyle\sum_{i=1}^{k}s_i - \displaystyle\sum_{j=k+1}^{n} s_j\right)\,,\quad\quad s_{i}=q\dfrac{\Delta\phi_i}{\phi_{\mathrm{msc}}}\,,
\end{equation}
with positions of the scattering centers coinciding with the locations of the ID hits from which the track is reconstructed. This was used to define a scattering significance for each track as the maximum of the individual scatterings, $S=\mathrm{max}\left\{|S(k)|,\,k=1,\,2,\,\cdots\,n\right\}$. These two variables were used to construct a composite variable,
\begin{equation}
C=\left|\dfrac{p_{\mathrm{loss}}}{p_{\mathrm{ID}}}\right|+rS\,,
\label{eq:composite_definition}
\end{equation}
where $r$ is a parameter that controls the relative contribution of each variable and was fixed at $r=0.07$ for optimal signal separation. The signal fraction, $f_{\mathrm{S}}$, was obtained by fitting the composite distribution obtained from the data, $\diff P/\diff C$, in each centrality and muon \pt\ bin to the form
\begin{equation}
\dfrac{\diff P}{\diff C}=f_{\mathrm{S}}\left. \dfrac{\diff P}{\diff C}\right|_{\mathrm{S}}+(1-f_{\mathrm{S}})\left. \dfrac{\diff P}{\diff C}\right|_{\mathrm{B}}\,,
\end{equation}
where $\left.\diff P/\diff C\right|_{\mathrm{S}}$ and $\left.\diff P/\diff C\right|_{\mathrm{B}}$ are the composite distributions obtained from the MC for the signal and background contributions respectively, and were obtained by constructing template probability distribution functions using a kernel estimation method included in the \verb=RooFit= package~\cite{Verkerke:2003ir}. Template fits are shown in Fig.~\ref{fig:muon_templates} for the $5<\pt<6$~\GeV\ (top) and $10<\pt<14$~\GeV\ (bottom) in the 0--10\% (left) and 60--80\% (right) centrality bins. In this analysis, muons were required to be on the interval $|\eta| < 1.05$. 
The extracted signal fraction was used to obtain the signal muon spectrum from the total muon yield, $N_{\mathrm{muon}}$, and ratios were constructed to form the \Rcp,
\begin{equation}
\Rcp(\pT)\Big|_{\mathrm{cent}} =
\dfrac{1}{\Rcollcent}
\dfrac{
\left.\dfrac{f_{\mathrm{S}}N_{\mathrm{muon}}}{\varepsilon\Nevt}\,\right|_{\mathrm{cent}} 
}{
\left.\dfrac{f_{\mathrm{S}}N_{\mathrm{muon}}}{\varepsilon\Nevt}\,\right|_{\mathrm{periph}} 
}\,.
\end{equation}
where $\varepsilon$ is the \pt- and centrality-dependent muon reconstruction efficiency obtained from the MC sample. The resulting \Rcp\ values are shown in the right panel of Fig.~\ref{fig:rcp_ratios}. The \Rcp\ values vary weakly with \pT\ and the points for each centrality interval are consistent with a \pT-independent \Rcp\ within the uncertainties on the points. The \Rcp\ varies strongly with centrality, increasing from approximately 0.4 in the 0-10\% centrality bin to approximately 0.85 in the 40-60\% bin.

\section{Conclusion}
Measurements of the centrality and \pt\ dependence of the suppression of single inclusive jets have been presented for jets reconstructed with the anti-\kt\ algorithm with $R=0.2,\,0.3,\,0.4$ and $0.5$. The results show that jets are suppressed in central collisions relative to peripheral collisions by approximately a factor of two. This suppression shows little variation with \pt\ over the range $38 < \pt < 210$~\GeV\ for all $R$ values. The effects of the suppression diminish in a gradual fashion in increasingly peripheral collisions. For jets with $\pt < 100$~\GeV, greater suppression is observed for jets with smaller $R$ values at the same \pt. Furthermore, the relative difference in suppression appears to be largest at smaller jet \pt\ which is qualitatively consistent with expected broadening due to radiative energy loss~\cite{He:2011pd}. The suppression of single muons from open heavy flavored hadrons was also measured and displays similar features in both the overall magnitude, centrality and \pt\ dependence of the suppression. This is different than the suppression observed for single hadrons at comparable \pt\ at the LHC which is rising steeply~\cite{CMS:2012aa,Aamodt:2010jd}. This feature is qualitatively different than the results from RHIC, where measurements of heavy quark suppression at RHIC via semi-leptonic decay channels in \AuAu\ collisions where comparable to the observed suppression of inclusive hadron for $\pt > 4$~\GeV\ \cite{Adare:2006nq,Abelev:2006db,Adare:2010de}.

\bibliographystyle{elsarticle-num}
\bibliography{HP2012Proceedings}






% \end{thebibliography}

\end{document}